\documentclass[	
						superscriptaddress,%
						twocolumn,%
						a4paper,
						showkeys,
					]{revtex4}

\usepackage[frenchb,english]{babel}
\usepackage[T1]{fontenc}
\usepackage{upgreek}
\usepackage{textcomp} 				
\usepackage{graphicx}			
\usepackage{epstopdf}				
\usepackage{color}					
\usepackage{multirow}				
\usepackage[nice]{nicefrac} 		
\usepackage{amssymb,amsmath}		
\usepackage{array}					
\usepackage{ulem}
\usepackage{float}
\usepackage{tabularx}
\usepackage{natbib}
\usepackage{longtable}

\definecolor{vert}{rgb}{0.5,0.758,0.5}
\definecolor{bleufonce}{rgb}{0,0,0.516}
\definecolor{orange}{rgb}{1,0.516,0}

\begin{document}

\title{Raman spectroscopy of rare-earth orthoferrites \textit{R}FeO$_3$ (\textit{R} = La, Sm, Eu, Gd, Tb, Dy)}
\date{\today}
\author{Mads Christof Weber}
\affiliation{Materials Research and Technology Department, Luxembourg Institute of Science and Technology, 41 rue du Brill, L-4422 Belvaux, Luxembourg}
\affiliation{Physics and Materials Science Research Unit, University of Luxembourg, 41 Rue du Brill, L-4422 Belvaux, Luxembourg}

\author{Mael Guennou}
\affiliation{Materials Research and Technology Department, Luxembourg Institute of Science and Technology, 41 rue du Brill, L-4422 Belvaux, Luxembourg}
\author{Hong Jian Zhao}
\affiliation{Materials Research and Technology Department, Luxembourg Institute of Science and Technology, 41 rue du Brill, L-4422 Belvaux, Luxembourg}
\author{Jorge Í$\tilde{\mbox{n}}$iguez}
\affiliation{Materials Research and Technology Department, Luxembourg Institute of Science and Technology, 41 rue du Brill, L-4422 Belvaux, Luxembourg}

\author{Rui Vilarinho}
\affiliation{IFIMUP and IN-Institute of Nanoscience and Nanotechnology, Physics and Astronomy Department of Faculty of Sciences of University of Porto, Porto, Portugal}
\author{Abílio Almeida}
\affiliation{IFIMUP and IN-Institute of Nanoscience and Nanotechnology, Physics and Astronomy Department of Faculty of Sciences of University of Porto, Porto, Portugal}
\author{Joaquim Agostinho Moreira}
\affiliation{IFIMUP and IN-Institute of Nanoscience and Nanotechnology, Physics and Astronomy Department of Faculty of Sciences of University of Porto, Porto, Portugal}
\author{Jens Kreisel}
\affiliation{Materials Research and Technology Department, Luxembourg Institute of Science and Technology, 41 rue du Brill, L-4422 Belvaux, Luxembourg}
\affiliation{Physics and Materials Science Research Unit, University of Luxembourg, 41 Rue du Brill, L-4422 Belvaux, Luxembourg}

\begin{abstract}
We report a Raman scattering study of six rare earth
orthoferrites \textit{R}FeO$_3$, with \textit{R}~= La, Sm, Eu, Gd, Tb,
Dy. The use of extensive polarized Raman scattering of SmFeO$_3$ and
first-principles calculations enable
the assignment of the observed phonon modes to vibrational symmetries
and atomic displacements. The assignment of the spectra and their
comparison throughout the whole series allows correlating the phonon
modes with the orthorhombic structural distortions of
\textit{R}FeO$_3$ perovskites. In particular, the positions of two
specific \textit{A}$_g$ modes scale linearly with the two FeO$_6$
  octahedra tilt angles, allowing the
distortion throughout the series. At variance with literature, we find that the two octahedra tilt angles scale
differently with the vibration frequencies of their respective
\textit{A}$_g$ modes. This behavior as well as the general relations
between the tilt angles, the frequencies of the associated modes and
the ionic radii are rationalized in a simple Landau model. The
reported Raman spectra and associated phonon-mode assignment provide
reference data for structural investigations of the whole series of
orthoferrites.
\end{abstract}


\maketitle

\section{Introduction}
In the past, \textit{R}FeO$_3$ perovskites have attracted considerable
interest due to their remarkable magnetic properties~\cite{White1969,
  White1982a, Eibschutz1967}. At ambient conditions, they adopt an
orthorhombic \textit{Pnma} structure, hence their common name
orthoferrites. This orthorhombic structure can be derived from the
ideal cubic perovskite structure by rotations (tilts) of its FeO$_6$
octahedra, where the tilt angles can be tuned by the size of the rare
earth \textit{R}. All members of the family possess a canted
antiferromagnetic structure arising from spin moments of the Fe$^{3+}$
cations. The antiferromagnetic ordering of the iron ions occurs
at a Néel temperature $T_\mathrm N$ around 650 to 700~K. Several
orthoferrites show a spin reorientation at lower temperatures. In
contrast to the Fe$^{3+}$ cations, the magnetic moments of the
\textit{R}$^{3+}$ rare earth ions order below 10~K. Interestingly, a
so-called compensation point where moments of the two sublattices
cancel has been reported for several \textit{R}FeO$_3$ compounds
\cite{White1969}. More recent studies have also focused on spin-ordering
processes of the rare earth ions \cite{Zhao2016, Cao2014,
  Marshall2012} and the interaction between magnetism and crystal
lattice, including the role of spin-lattice coupling in multiferroic
properties~\cite{Tokunaga2009,Du2010a,Lee2011,Kuo2014,Cheng2014a}.
 
Tilts of the FeO$_6$-octahedra are the main structural parameters to tune the band
overlap and thus the physical properties of orthoferrites. Unfortunately, tilt angles are chronically difficult to
probe directly, specifically in thin films, because they require in
depth diffraction experiments needing at best large-scale instruments
using neutron or synchrotron radiation. Alternatively, Raman
spectroscopy (RS) is a well-known technique to follow tilt-driven soft
mode phase transitions \cite{Fleury1968,Scott1969,Scott1974}. More
recently, it has been shown that RS is also an appropriate probe for
the investigation of lattice distortions and slight changes in
octahedra rotations \cite{Iliev2006a, Chaix-Pluchery2011, Weber2012,
  Todorov2012}. Further to this, RS is an ideal probe for the
investigation of spin-phonon coupling phenomena
\cite{Laverdiere2006,Ferreira2009,Moreira2010,Bhadram2013,ElAmrani2014}. Finally,
RS is a now widely used technique for probing even subtle
strain-induced structural modifications in oxide thin films
\cite{Tenne2006,Weber2016,Kreisel2012}. All such investigations rely on thorough
reference spectra, solid knowledge of the relations between structural
distortions phonon modes, and on a proper band assignment of
vibrational bands in terms of symmetry and atomic displacement
patterns. The present paper aims at providing this fundamental
knowledge by investigating both experimentally and theoretically a
series of orthoferrites and by proposing a consolidated view of this new
data together with available literature data on other
members of the family.

\section{Experimental}

SmFeO$_3$ single crystals were grown in an optical-floating-zone
furnace as described elsewhere~\cite{Cao2014}. Three single domain
platelets were oriented along the three orthorhombic directions, with
their edges also parallel to cristallographic axes, and polished down
to a thickness of 100 $\mathrm{\mu}$m. The single domain state was
verified by XRD and polarized light microscopy. A SmFeO$_3$ crystal
was manually grinded to acquire a homogeneous powder. LaFeO$_3$ and
EuFeO$_3$ powders were obtained by conventional solid state
reactions. GdFeO$_3$ and DyFeO$_3$ powder samples were prepared using
the urea sol-gel combustion method, reported elsewhere
\cite{Moreira2010a} and their quality was checked by XRD and
SEM. TbFeO$_3$ samples were prepared by floating zone method in
FZ-T-4000 (Crystal Systems Corporation) mirror furnace. As starting
materials, Fe$_2$O$_3$ (purity 2N, supplier: Sigma Aldrich), and
Tb$_4$O$_7$ (purity 3N, supplier: Alpha Aesar) were used. They
were mixed in a Tb:Fe stoichiometric ratio, cold pressed into rods and
sintered at 1100 $^{\circ}$C from 12 to 14 hours in air. Their quality
was checked by X-ray powder diffraction and by energy dispersion X-ray
analysis, confirming the single perovskite phase.\\

Raman scattering measurements were performed with an inVia Renishaw
Reflex Raman Microscope in micro-Raman mode. For excitation a 633~nm
He-Ne laser with a spectral cut-off at 70~cm$^{-1}$ was used. Great
care was taken to avoid heating of the sample by limiting the incident laser power. Samples were
cooled to liquid nitrogen temperature in a Linkam THMS600 stage in
order to reduce thermal broadening of the spectra and ease the
identification of Raman bands. The band positions were obtained
by fitting the spectra with Lorentzian functions.\\

For the calculations we used density functional theory (DFT)
within the generalized gradient approximation revised for
solids,\cite{Perdew2008} as implemented in the Vienna ab-initio
Simulation Package ({\sc VASP})\cite{Kresse1996,Kresse1999}. For a
better treatment of iron's 3$d$ electrons, we used the Hubbard-like
correction proposed by Dudarev {\sl et al}.,\cite{Dudarev1998} with
$U_{\rm eff}$~=~4~eV. The ionic cores were treated within the
projection augmented approximation (PAW),\cite{blochl94} and the
following electrons were explicitly solved in the simulations: O's
2s$^2$2p$^4$; Fe's 3p$^6$3d$^7$4s$^1$; 5p$^6$5d$^1$6s$^2$ for Eu,
Gd, Tb, and Dy; and 5s$^2$5p$^6$5d$^1$6s$^2$ for Sm and La. Note
that, for the generation of the PAW potentials of the rare-earth
species, a $3+$ ionization state was assumed and the remaining 4$f$
electrons were considered to be frozen in the ionic core. We
explicitly checked in one case (GdFeO$_{3}$) that this approximation
has a very small impact on the phonon frequencies and eigenvectors
of interest in this work. Electronic wave functions are described in
a basis of plane waves cut-off at 500~eV; reciprocal space integrals
in the Brillouin zone of the 20-atom $Pnma$ cell were computed in a
mesh of $4\times3\times5$ $k$-points. Structural optimization were
performed until residual atomic forces are smaller than
0.01~eV/\AA\, and phonon spectra were computed calculated by the
finite difference method.

\section{Results and Discussion}
\subsection{Structural properties of the \textit{R}FeO$_3$ series}

Rare earth orthoferrites crystallize in an orthorhombic \textit{Pnma}
structure at ambient conditions.  With respect to the parent cubic
perovskite phase \textit{Pm$\overline{3}$m}, the \textit{Pnma} structure in orthoferrites can be derived by octahedral
rotations. In Glazer's notation the octahedra tilt system is expressed
as \textit{a$^-$b$^+$a$^-$} \cite{Glazer1972} or in pseudo-cubic
settings as rotations $\theta$, $\phi$ and $\Phi$ around the
[101]$_{\mathrm{pc}}$, [010]$_{\mathrm{pc}}$ and
[111]$_{\mathrm{pc}}$, respectively \cite{Mitchell2002}. Megaw has
shown that it is sufficient to consider two independent angles
$\theta$ and $\phi$ in order to describe the octahedral rotations of
the \textit{Pnma} phase, assuming that the octahedral tilts
\textit{a$^-_x$} and \textit{a$^-_z$} are approximately equal
\cite{Megaw1973}. The angle $\Phi$ can be then expressed as
$\cos\Phi=\cos\theta \cos\phi$ \cite{Mitchell2002}. The
octahedra rotations represent the order parameters for a hypothetical
phase transition to the cubic \textit{Pm$\overline{3}$m} phase.  \\

\begin{figure}[]	
	\begin{center}
		\includegraphics[]{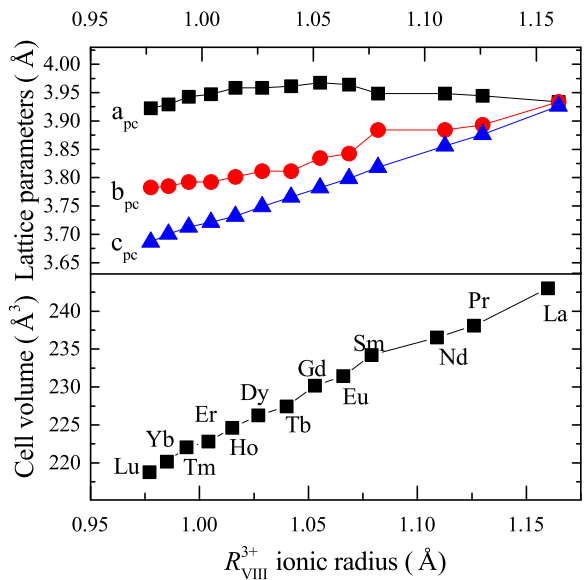}
		\caption{Variation of the pseudocubic cell parameters and orthoromic unit-cell volume as a function of the \textit{R$^{3+}_{VIII}$} ionic radius \cite{Shannon1976}. The lattice parameters are taken from ref.\cite{Marezio1970,Marezio1971}.}
		\label{fig:Structure}
	\end{center}
\end{figure}

Similar to other perovskites with \textit{Pnma} structure, such
as orthochromites \textit{R}CrO$_3$, orthomanganites
\textit{R}MnO$_3$, orthonickelates \textit{R}NiO$_3$, or
orthoscandates \textit{R}ScO$_3$, we can assume in good approximation
that changing the rare earth affects negligibly the chemical bonding
of the material. In contrast, the size of the rare earth impacts on
the distortions of the structure, as measured for example by the tilt
angles or the spontaneous strains, and can be continuously tuned by
the size of the \textit{R}$^{3+}$ rare earth. The octahedral rotations
are most reliably calculated from atomic positions following the
formalism in Ref.~\onlinecite{Zhao1993}. Table~\ref{table_1}
summarizes the structural properties of all members of the
\textit{R}FeO$_3$ family. In Fig.~\ref{fig:Structure} the structural
evolution throughout the series is illustrated by the unit cell
volume, which scales linearly with the ionic radius, and the lattice
parameters. \\

The tolerance factor, given in Table~\ref{table_1}, is an indication
for the stability of the perovskite structure. The closer its value is
to $1$, the closer the structure is to the cubic structure. From
both, pseudo-cubic lattice parameters and tolerance factor, we find
that with increasing ionic radius of the rare earth, from lutetium to
lanthanum, the structure approaches a cubic metric. Notably, LaFeO$_3$
appears to be closest to a cubic structure.

\begin{table*}[th!]
	\centering
	\caption{Structural characteristics of \textit{R}FeO$_3$ samples: \textit{R$^{3+}$} ionic radii (r$_{R^{3+}}$ values given in an eightfold environment~\cite{Shannon1976}, lattice parameters, tolerance factor \textit{t} calculated from the ionic radii following \cite{Mitchell2002}: $t=(r_{R^{3+}}+r_{O^{2-}})/(\sqrt{2}(r_{Fe^{3+}}+r_{O^{2-}}))$ and octahedra tilt angles ($\phi$[010], $\theta$[101]) calculated from the atomic coordinates. The data for the crystal structures are from Refs.~\cite{Marezio1970,Marezio1971}.}
	\label{table_1}
	\setlength{\extrarowheight}{4pt} 
	\begin{tabularx}{\textwidth}{XXXXXXXcXX}
		\hline\hline
		\multicolumn{1}{r}{} & \multicolumn{1}{r}{} & \multicolumn{3}{c}{Lattice parameters ($Pnma$ setting)}    & \multicolumn{1}{r}{} & \multicolumn{1}{r}{} & \multicolumn{1}{c}{} & \multicolumn{2}{c}{FeO$_6$ octhahedra tilt angle} \\ \cline{2-7} \cline{9-10}
		& r$_{R^{3+}}$  & a (\r{A}) & b (\r{A}) & c (\r{A}) & V (\r{A}$^3$)  & t     &                      & $\phi$ {[}010{]} ($^\circ$)                              & $\theta$ {[}101{]} ($^\circ$)                              \\ \cline{2-10}
		LaFeO$_3$               & 1.160 & 5.563 & 7.867 & 5.553 & 243.022 & 0.934 &                      & 7.3	                                     & 12.2                                      \\
		PrFeO$_3$               & 1.126 & 5.578 & 7.786 & 5.482 & 238.085 & 0.921 &                      & 9.6	                                     & 13.6                                      \\
		NdFeO$_3$               & 1.109 & 5.584 & 7.768 & 5.453 & 236.532 & 0.915 &                      & 10.0	                                     & 14.5                                      \\
		SmFeO$_3$               & 1.079 & 5.584 & 7.768 & 5.400 & 234.233 & 0.904 &                      & 11.2                                      & 15.6                                      \\
		EuFeO$_3$               & 1.066 & 5.606 & 7.685 & 5.372 & 231.437 & 0.899 &                      & 11.6                                      & 16.0                                      \\
		GdFeO$_3$               & 1.053 & 5.611 & 7.669 & 5.349 & 230.172 & 0.894 &                      & 11.9                                      & 16.2                                      \\
		TbFeO$_3$               & 1.040 & 5.602 & 7.623 & 5.326 & 227.442 & 0.889 &                      & 12.1                                      & 16.9                                      \\
		DyFeO$_3$               & 1.027 & 5.598 & 7.623 & 5.302 & 226.255 & 0.884 &                      & 12.6                                      & 17.3                                      \\
		HoFeO$_3$               & 1.015 & 5.598 & 7.602 & 5.278 & 224.611 & 0.880 &                      & 12.7                                      & 17.7                                      \\
		ErFeO$_3$               & 1.004 & 5.582 & 7.584 & 5.263 & 222.803 & 0.876 &                      & 12.9                                      & 18.2                                      \\
		TmFeO$_3$               & 0.994 & 5.576 & 7.584 & 5.251 & 222.056 & 0.872 &                      & 12.9                                      & 18.6                                      \\
		YbFeO$_3$               & 0.985 & 5.557 & 7.570 & 5.233 & 220.134 & 0.869 &                      & 13.4                                      & 19.0                                      \\
		LuFeO$_3$               & 0.977 & 5.547 & 7.565 & 5.213 & 218.753 & 0.866 &                      & 13.2                                      & 19.5                                     
	\end{tabularx}
\end{table*}

\subsection{Raman spectra and mode assignment}

\begin{figure}[]
	\begin{center}
		\includegraphics{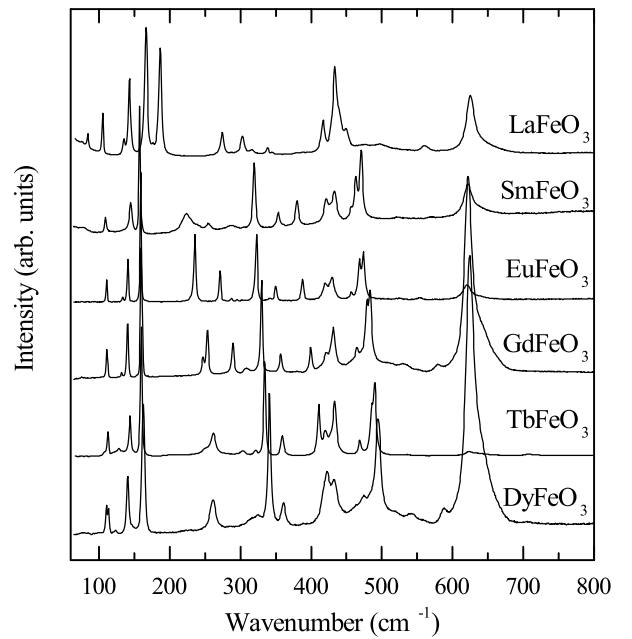}
		\caption{Raman spectra at 80~K of six rare earth orthoferrites \textit{R}FeO$_3$	(\textit{R} = La, Sm, Eu, Gd, Tb, Dy) collected with a 633~nm He-Ne laser line.}
		\label{fig:spectra}
	\end{center}
\end{figure}

The orthorhombic \textit{Pnma} structure gives rise to 24 Raman-active
vibrational modes \cite{Kroumova2003}, which decompose into
$\Gamma_{\mathrm{Raman}}$~$=$ 7\textit{A}$_g$ + 5\textit{B}$_{1g}$ +
7\textit{B}$_{2g}$ + 5\textit{B}$_{3g}$. Schematically, the vibration
modes below 200~cm$^{-1}$ are mainly characterized by displacements of
the heavy rare-earth ions. Above 300~cm$^{-1}$, motions of the light
oxygen ions dominate, and in the intermediate frequency range
vibration patterns involve both ions. Note that iron ions 
occupy centers of inversion in the \textit{Pnma} structure and,
therefore, vibrations involving Fe$^{3+}$ motions are not Raman-active.  Fig.~\ref{fig:spectra} shows the Raman spectra of six rare
earth orthoferrites \textit{R}FeO$_3$ (\textit{R}~= La, Sm, Eu, Gd,
Tb, Dy), all measured at 80~K in order to reduce thermal broadening
and make mode identification easier. Thanks to well-defined spectra,
we identify between 18 and 21 vibration bands, depending on the
compound. The remaining predicted modes are either masked by band
overlap or their intensity is below the detection limit. The Raman
spectra of SmFeO$_3$, EuFeO$_3$, GdFeO$_3$, TbFeO$_3$ and DyFeO$_3$
present a similar overall shape which allows to follow the
evolution of particular bands throughout the series. The spectral
signature of LaFeO$_3$ is distinctly different as explained by the
size difference between La$^{3+}$ and the closest Sm$^{3+}$ and also
its proximity to the cubic structure (see
Fig.~\ref{fig:Structure}). This is similar to observations for other
rare earth perovskites, where the Raman spectrum of the lanthanum
member is systematically different when compared to the remaining
members of the series \cite{Iliev2006a, Weber2012, Daniels2013}. This
will be elaborated on later in the discussion.

\begin{figure}[]	
	\begin{center}
		\includegraphics{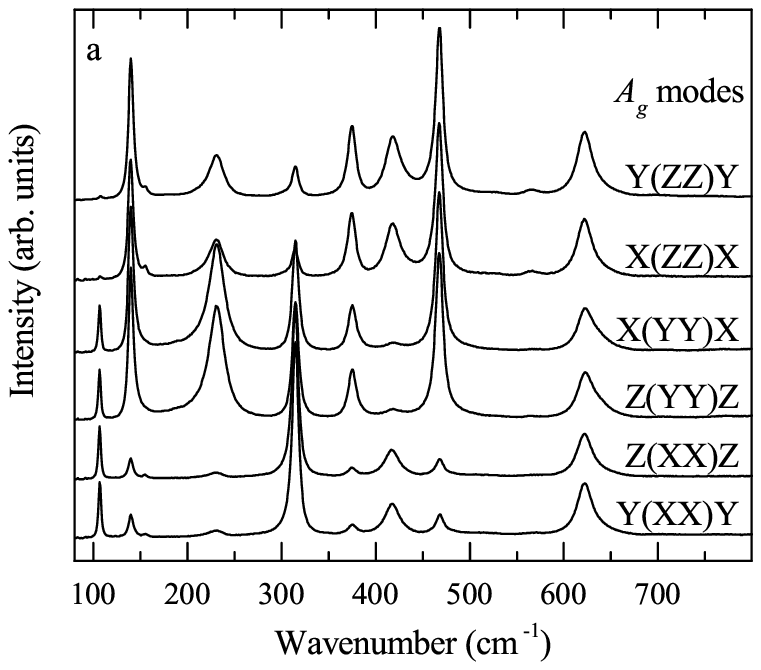}
		\includegraphics{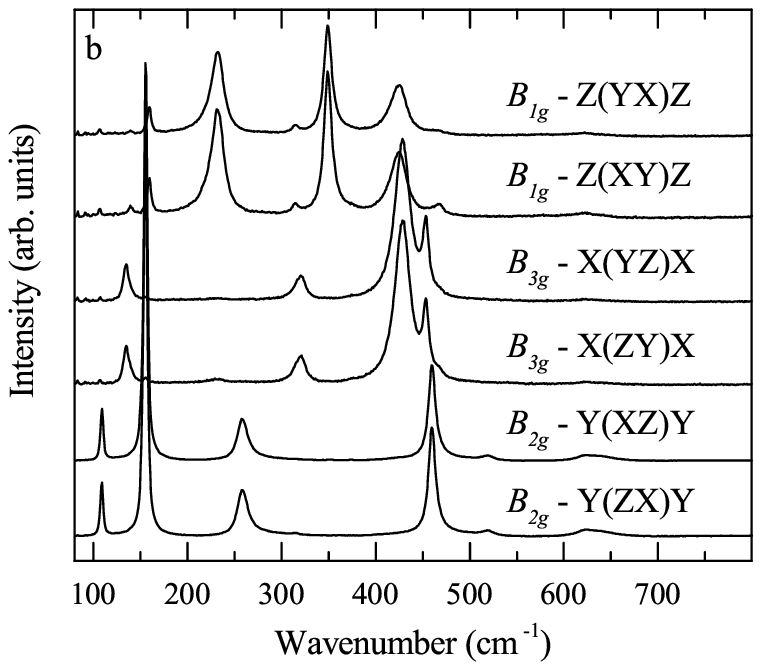}
		\caption{Polarized Raman spectra of SmFeO$_3$ at ambient conditions. The measurement configurations are given in Porto's notation, in a) for the vibration modes of \textit{A$_g$} symmetry and in b) for the vibration modes of \textit{B$_{1g}$}, \textit{B$_{2g}$} and \textit{B$_{3g}$} symmetries. X, Y and Z correspond to the orthorhombic axes in the \textit{Pnma} space group.}
		\label{fig:SmFeO3modes}
	\end{center}
\end{figure}

In order to go further in the mode assignment, we performed a
polarized Raman study of SmFeO$_3$ single crystals. Indeed, the
identification of the symmetry of the difference Raman bands is
difficult if not impossible from powder samples alone. On the other
hand, polarized Raman spectroscopy on well-oriented single crystals
allows identifying the symmetries of phonon bands. Raman modes of a
given symmetry can be selectively probed through particular configurations
of incident and scattered light polarizations with respect to the
orientation of the crystal. This experimental configuration is
expressed in Porto's notation~\cite{Damen1966}. In the following, we
use X, Y, Z to indicate the crystallographic axes in the
  \textit{Pnma} setting. Figure~\ref{fig:SmFeO3modes} presents the
obtained results for SmFeO$_3$ single crystals for twelve scattering
configurations. Fig.~\ref{fig:SmFeO3modes}a shows the Raman spectra
for \textit{A}$_g$ configurations modes, while spectra exhibiting
\textit{B}$_{1g}$, \textit{B}$_{2g}$ or \textit{B}$_{3g}$ modes are
given in Fig.~\ref{fig:SmFeO3modes}b. In total, we identify all
expected \textit{A}$_g$, six \textit{B}$_{2g}$ modes and four out of
five \textit{B}$_{1g}$ and \textit{B}$_{3g}$ modes (see
Table~\ref{table_2}).

In a next step, we run DFT calculations of phonon modes for all
measured orthoferrites in order to confirm the mode symmetries and
associate a vibrational pattern to each mode. A summary of all
theoretical and experimental band frequencies with their symmetry and
characteristic atomic motions is given in
Table~\ref{table_2}. The calculated frequencies are in very good
agreement with our experimental values and the continuous evolution of
the spectral signature.\\

The band between 600 and 650~cm$^{-1}$ in Fig.~\ref{fig:SmFeO3modes}a
shows a peculiar behaviour and needs a specific discussion. First, as
can be seen in Fig.~\ref{fig:spectra}, its frequency seems to be
independent of the rare earth. Besides, it shows intensity variations
from sample to sample that contrast with the other bands, and also
exhibits a strong asymmetry. For SmFeO$_3$, Fig.~\ref{fig:SmFeO3modes}
shows that these bands appear with very low intensity in crossed
polarization but are strongly visible in parallel configuration, which
would rather point to a \textit{A}$_g$ symmetry. However, as can be
seen in Table~\ref{table_2}, the calculations predict two bands of
\textit{B}$_{2g}$ and \textit{B}$_{3g}$ symmetry in this region, but
no \textit{A}$_g$ Raman mode, and all \textit{A}$_g$ modes are already
conclusively attributed. We therefore conclude that this band is not a
first-order Raman mode.

A precise interpretation for this band is beyond the scope of this work, but we note that similar features have been described for other perovskite oxides, with unclear
assignments and conflicting reports. As an example, Iliev et
al. discussed it for LaCrO$_3$ \cite{Iliev2006c} and demonstrated that
its intensity can be reduced by annealing the sample in
vacuum. Therefore it seems likely that it is related to chemical
defects of the lattice \cite{Iliev2006c}. Here, we note that DyFeO$_3$
and GdFeO$_3$, where this band is stronger, were produced by a
chemical metalorganic process, whereas the other samples (LaFeO$_3$,
SmFeO$_3$, EuFeO$_3$ and TbFeO$_3$) were synthesized by solid-state
reaction. A difference in defect chemistry originating from different
growth processes is therefore plausible.

\squeezetable 
\begin{table*}[!ht]
	\centering
	\caption{Experimental and theoretical band positions, the corresponding symmetry assignment and main atomic motions of the observed Raman modes in \textit{R}FeO$_3$. The main atomic motions are extracted from DFT calculations. Refer to the supplemental materials for the displacements of all ions.}
	\label{table_2}
	
	\setlength{\extrarowheight}{4pt} 
	
	\begin{tabularx}{\textwidth}{c >{\centering}p{0.9cm} >{\centering}p{0.9cm} >{\centering}p{0.9cm} >{\centering}p{0.9cm} >{\centering}p{0.9cm} >{\centering}p{0.9cm} >{\centering}p{0.9cm} >{\centering}p{0.9cm} >{\centering}p{0.9cm} >{\centering}p{0.9cm} >{\centering}p{0.9cm} >{\centering}p{0.9cm} X}
		\hline\hline
		\multicolumn{1}{c}{Symmetry} & 
		\multicolumn{2}{c}{LaFeO$_3$} & 
		\multicolumn{2}{c}{SmFeO$_3$} & 
		\multicolumn{2}{c}{EuFeO$_3$} & 
		\multicolumn{2}{c}{GdFeO$_3$} & 
		\multicolumn{2}{c}{TbFeO$_3$} & 
		\multicolumn{2}{c}{DyFeO$_3$} & 
		\multicolumn{1}{c}{main atomic motion}
		
		\\
		& 80 K & calc. & 80 K & calc. & 80 K & calc. & 80 K & calc. & 80 K & calc. & 80 K & calc. & 
		\\
		\hline
		A$_g$(1) & 84.5  &89 & 109.5 & 109 & 110.9 & 112 & 111.1 & 111 & 112.5 & 112 & 113.3 & 112 & $R$(x), in-phase in x-z, out-of-phase in y 
		\\
		A$_g$(2) & 135.3  &127 & 144.2 &138	& 140.7 & 140 & 140.4 & 137 &		143.9 & 136 &	140.5 & 135 & $R$(z), out-of-phase
		\\
		A$_g$(3) & 186.6  &183 & 223.9	& 244 & 235.1	& 252 & 253.2 & 255 &	261.9 & 259 &	261.5 & 262 & [010]$_{\mathrm{pc}}$ FeO$_6$ rotation, in-phase
		\\
		A$_g$(4) & 274.2 & 273 & 319.1 & 320 &	323.2 & 325 & 329.9 & 330&	334.5 & 330 &	341.1 & 332 & O(1) x-z plane
		\\
		A$_g$(5) & 302.8  &306 & 379.5 & 383	& 387.8 & 397 & 399.4 &	405 &	410.9 & 410 & 422.4 & 422 &	[101]$_{\mathrm{pc}}$ FeO$_6$ rotation, in-phase
		\\
		A$_g$(6) & 449.8 & 433 & 420.7 & 413 &	419.6 & 414 & 420.9 &	416 & 420.1 & 416 & 417.3 & 415 & Fe-O(2) stretching, in-phase
		\\
		A$_g$(7) &  433.3 & 413 & 470.7 & 468 & 474.0 & 476 & 483.6 & 480 &	490.1 & 484 & 496.8 & 490 & O(1)-Fe-O(2) scissor-like bending
		\\
		B$_{1g}$(1) &       & 169 & 160.7 & 151 &       & 149 &       & 143 &       & 139 &       & 135 &	$R$(y) in-phase in x-z, out-of-phase in y
		\\
		B$_{1g}$(2) &       & 148 & 238.7	& 233	& 236.4 & 243 & 247.1 & 244 & 251.9 & 248 &       & 250 & [010]$_{\mathrm{pc}}$ FeO$_6$ rotation, out-of-phase
		\\
		B$_{1g}$(3) & 338.1 & 328 & 353.3	& 352 & 350.0 & 356 & 357.0 & 356	& 359.2 & 356 & 360.9 & 359 & [010]$_{\mathrm{pc}}$ FeO$_6$ rotating, out-of-phase
		\\
		B$_{1g}$(4) & 442.3 & 425 & 426.4	& 422	& 425.8 & 424 & 428.8 & 426 & 427.7 & 425 & 427.4 & 427 & Fe-O(2) stretching, out-of-phase
		\\
		B$_{1g}$(5) & 560.9 & 584 &       & 594 &       & 597 &       & 595 &       & 592 &       & 593 & Fe-O(1) stretching
		\\
		B$_{2g}$(1) & 105.5 & 103 & 109.8 & 109 & 110.9 & 111 & 111.1 & 109 & 107.7 & 109 & 110.6 & 109 & $R$ (z), in-phase in x-z, out-of-phase in y
		\\
		B$_{2g}$(2) & 143.0 & 144 & 157.4 & 159	& 159.3 & 163 & 159.9 & 161 & 160.1 & 161 & 162.8 & 161 & $R$ (x), out-of-phase
		\\
		B$_{2g}$(3) & 166.5 & 172 & 255.0 & 278 & 271.1 & 291 & 289.3 & 299 & 302.7 & 305 & 324.9 & 311	& [101]$_{\mathrm{pc}}$ FeO$_6$ rotation, in-phase
		\\
		B$_{2g}$(4) &       & 329 &       & 346 &       & 348 &       &	349 &       & 349 &       & 351	& O(1) x-z plane
		\\
		B$_{2g}$(5) & 416.8 & 401 & 462.8 & 460 & 468.2 & 469 & 478.9 & 474 & 485.6 & 478 & 493.7 & 482 & O(1)-Fe-O(2) scissor-like bending
		\\
		B$_{2g}$(6) &       & 481 & 521.5 & 513 & 524.5 & 521	& 531.7 & 528 & 535.8 & 528 &	      & 534	&	O(2)-Fe-O(2) scissor-like bending, in-phase
		\\
		B$_{2g}$(7) & 625.1 & 622 & 640.5 & 610 & 638.1 & 613 & 640.5 & 612 &       & 611 & 624.2 & 612	& Fe-O(2) stretching, in-phase
		\\
		B$_{3g}$(1) &       & 137 & 145.0 & 135 & 133.6 & 134	& 132.2 & 129 &       & 126 &       & 123 & $R$ (y) out-of-phase in x-z, y
		\\
		B$_{3g}$(2) & 316.8 & 300 & 322.8 & 313 &       & 315 &       & 312 &       & 311 &       & 311	& O(1)-Fe-O(2) in-phase 
		\\
		B$_{3g}$(3) & 436.0 & 425 & 432.7 & 424 & 429.9 & 424 & 431.5 & 426 & 433.3 & 422 & 433.1 & 424 & octahedra squeezing in y
		\\
		B$_{3g}$(4) & 428.6 & 408 & 455.9 & 447 & 456.7 & 452 & 465.0 & 455 & 468.8 & 457 & 473.7 & 460 & O(2)-Fe-O(2) scissor-like bending, out-of-phase
		\\
		B$_{3g}$(5) & 641.9 & 650 &       & 641 &       & 643 &       &	640 &       & 637	& 639.4 & 637	& FeO$_6$ breathing 
		
	\end{tabularx}
\end{table*}

\subsection{Phonon Raman modes vs. ionic radii and octahedra tilt angle}

\begin{figure}[]
 	\begin{center}
 		\includegraphics{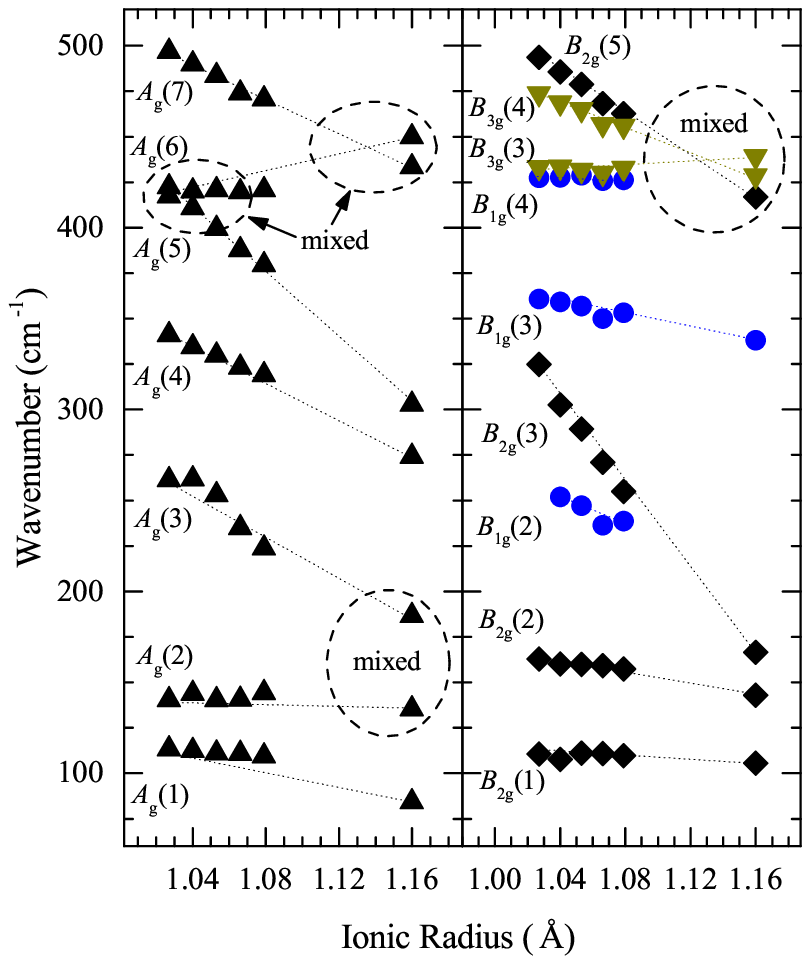}
 		\caption{Raman phonon wavenumbers of \textit{R}FeO$_3$
 			as a function of the rare earth \textit{R$^{3+}$} ionic radius. All lines are guides to the eye only.}
 		\label{fig:Allwave}
 	\end{center}
\end{figure}

Figure~\ref{fig:Allwave} presents the evolution of the band positions
for the different orthoferrites as a function of the ionic radii of
the rare earths. Overall, Raman bands shift to lower frequencies
with increasing r$_{R^{3+}}$, which naturally correlates with the
increase in volume and, therefore, of most bond lengths. It can
be seen that the frequencies of the Raman modes are differently
sensitive to the change of rare earth. This is understood in the context of the structural instabilities in the \textit{Pnma}
structure. In the framework of Landau theory, the two octahedra
  rotations represent the two order parameters for a phase transition
from the high-symmetry parent cubic perovskite phase. If a
vibrational displacement is directly related to the order parameter,
the phonon mode is called soft mode \cite{Hayes1978, Scott1974} and
can experience very large frequency shifts. \\

Thus our Raman data across the {\sl R}FeO$_{3}$ family exhibits
patterns that provide useful insights into the relations among
structural order parameters, associated phonon frequencies, and steric
effects driven by the {\sl R} cation. To understand such patterns
better, it is useful to think in terms of the simplest Landau-like
potential connecting all the relevant ingredients, which we introduce
in the following. Let $Q$ denote the relevant structural order
parameter, which may correspond to either antiphase or in-phase
FeO$_{6}$ rotations in the case of orthoferrite perovskites. Also, let
$\eta$ be the isotropic strain of the material, and let us assume that
$Q$~=~$\eta$~=~0 corresponds to the ideal cubic perovskite. We can
write the Landau free energy as a function of these variables as
\begin{equation}
	\Delta F(Q,\eta) = \frac{1}{2} A' (T-T_{\rm t}) Q^{2} + \frac{1}{4} B
	Q^{4} + \frac{1}{2} C \eta^{2} + \gamma \eta Q^{2} \; ,
\end{equation}
where the $A'(T-T_{\rm t})$ and $B$ parameters define the potential
well associated to the $Q$-instability, and we have assumed the
simplest temperature ($T$) dependence of the quadratic term as
customarily done in Landau theory. We want to focus on the behavior of
the material at temperatures well below the structural transition
between the cubic and orthorhombic phases; hence, the transition
region is of no interest to us and we can assume a simple fourth-order
potential to describe the energy surface, introducing a characteristic
$T_{\rm t} \gg T$ whose precise meaning (i.e., whether or not it
coincides with the actual transition temperature) is irrelevant
here. Our Landau-like potential also includes an elastic
constant $C$ that quantifies the stiffness of the material, as well as
the lowest-order coupling between $\eta$ and $Q$ that is allowed by
symmetry. (Because $\eta$ is a fully-symmetric strain, the
  coupling that goes as $\sim \eta Q^{2}$ always exists irrespective
of the symmetry of $Q$; further, this is the lowest-order coupling
provided that $Q$ is a symmetry-breaking order parameter, as it is the
case here.)

In principle, we could write such a potential for each of the
orthoferrites, fit the corresponding parameters to reproduce
experimental data, etc. However, here we would like to test the
following hypothesis: we assume that all the {\sl R}FeO$_{3}$
orthoferrites present the same parameters quantifying the energetics
of $Q$ and $\eta$, and that the only feature changing from compound to
compound is the value of the strain $\eta$, as dictated by the size of
the rare-earth cation. More specifically, let $\bar{r}$ be a reference
value for the ionic radii of the ${\sl R}^{3+}$ cations (for the sake
of concreteness, we can think of $\bar{r}$ as an average value),
and let $r$ be the radius of the rare-earth cation for a particular
{\sl R}FeO$_{3}$ compound; then, such a compound is characterized by a
strain $\eta = \kappa (r-\bar{r})$, where $\kappa$ is a suitable
proportionality constant. We can further consider a Landau potential
as the above one, but corresponding to some sort of average
orthoferrite (i.e., with parameters obtained as an average of the
parameters of specific compounds), substitute the expression for $\eta
= \eta (r)$, and postulate the resulting $r$-dependent potential as
applicable to the whole family:
\begin{equation}
	\Delta F(Q,r) = \frac{1}{2}[A'(T-T_{\rm
		t})+\kappa\gamma(r-\bar{r})]Q^{2} + \frac{1}{4}BQ^{4} \; .
\end{equation}
A key point to realize here is that the compound dependence is
restricted to the harmonic part of the potential. Further, formally,
the ionic radius $r$ plays the exact same role as the temperature.

Now, let us introduce $\bar{A} = A' (T-T_{\rm t}) +
\kappa\gamma(r-\bar{r})$. Then, it is straightforward to derive
\begin{equation}
	Q_{\rm eq} = \pm (-\bar{A}/B)^{1/2} \; 
\end{equation}
for the equilibrium order parameter at $T \ll T_{\rm t}$. (At such a
temperature, we assume $\bar{A}<0$ for all relevant $r$ values.)
Further, the associated soft-mode frequency is
\begin{equation}
	\omega = \sqrt{\frac{2B}{m}}\,Q_{\rm eq} \; ,
\end{equation}
where $m$ is a mass characteristic of the $Q$ order parameter; for our
FeO$_{6}$-rotational cases, this reduces to the mass of the oxygen
atom. Now, combining these equations we can write
\begin{equation}\label{relation}
	\omega = \sqrt{\frac{2B}{m}}\,Q_{\rm eq} = \left[-\frac{A'(T-T_{\rm t}) + \kappa\gamma (r - \bar{r})}{m/2}\right]^{1/2} \; ,
\end{equation}
which gives us the desired relation between the compound-dependent
parameter (the ionic radius $r$), temperature, the relevant structural
distortion, and its corresponding phonon frequency. In other words, we
expect a linear relation between order parameter $Q$, i.e. the octahedra tilt angle, and the corresponding soft-mode frequency
$\omega$, which is solely dependent on the ionic radius $r$ of the
rare earth and the temperature $T$. In particular, if we fix $T =
T_{\rm RT}$, this expression allows us to compare (and predicts the
behavior of) the structural and Raman data across the orthoferrite
series.

In order to apply this relation to the orthoferrite family, the
identification of the soft modes is crucial. An order parameter may
give rise to several soft modes which do not necessarily need to be
Raman active. However, using the group theoretical formalism of Landau
theory, Birman \cite{Birman1973a} and Shigenari \cite{Shigenari1973}
demonstrated that one of the soft modes related to an order parameter
has a Raman-active $A_g$ symmetry. In the \textit{Pnma} structure it
is therefore common to focus on the $A_g$ soft modes. From our DFT
calculations leading to the assignment of the bands to the respective
vibrational pattern (see Table~\ref{table_2}), we find that $A_g$(3)
and $A_g$(5) are the soft-modes to corresponding
$Q_{[010]_{\mathrm{pc}}}$ and $Q_{[101]_{\mathrm{pc}}}$, respectively,
where $Q_{[010]_{\mathrm{pc}}}$ and $Q_{[101]_{\mathrm{pc}}}$ are the
order parameters of the \textit{Pnma} structure representing the
octahedra rotations around the [010]$_{\mathrm{pc}}$ (in-phase)
and the [101]$_{\mathrm{pc}}$ (antiphase) axes. The assignment
of the $A_g$(3) as a soft-mode is at variance with earlier work by
Todorov and co-workers \cite{Todorov2012} and underlines the
importance of precise calculations to gain full understanding of the
experimental findings.

\begin{figure}[]
	\begin{center}
		\includegraphics{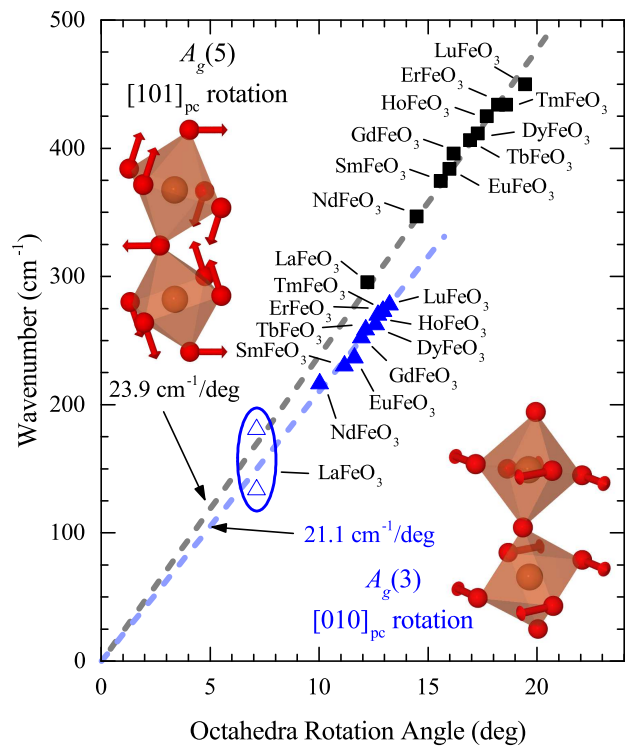}
		\caption{\textit{R}FeO$_3$ soft-mode wavenumbers at ambient conditions as a function of octahedra-tilt angles for the \textit{A$_g$}(3) and \textit{A$_g$}(5) modes. Data from this work are complemented by literature data for NdFeO$_3$~\cite{Singh2008}), HoFeO$_3$ and TmFeO$_3$~\cite{Venugopalan1985}, ErFeO$_3$~\cite{Koshizuka1980}) and LuFeO$_3$~\cite{Venugopalan1990}. The oval and the open symbols point out the mode mixing region of the $A_g$(2) and $A_g$(3) modes for LaFeO$_3$, as discussed in the text.}
		\label{fig:TiltFrequ}
	\end{center}
\end{figure}

Fig.~\ref{fig:TiltFrequ} presents the evolution of the soft modes
$A_g$(3) and $A_g$(5) against the corresponding tilt angle. For
completeness and in order to test the general validity of this model,
we extend our graph with literature data on orthoferrites with Lu, Tm,
Er, Ho and Nd \cite{Venugopalan1990, Venugopalan1985, Koshizuka1980,
  Singh2008}. The evolution shows the expected linear relation between
the vibrational frequencies and the tilt angles of the
\textit{R}FeO$_3$. This adds further support to the proposed
soft-mode-like relation of tilt frequency and size of the rare earth,
not only in the orthoferrites, but also for other families where this
behavior has been experimentally verified: orthomanganites
\cite{Iliev2006a}, orthochromates \cite{Weber2012}, orthoscandates
\cite{Chaix-Pluchery2011} among others \cite{Todorov2012}. However, at
variance with these previous experimental data, our work on
orthoferrites show two additional features that have to be commented
on, namely that i) the two tilt modes follow two different lines and
ii) LaFeO$_3$ deviates significantly from the general linear
behaviour.

The octahedral-rotation angles and soft-mode frequencies do not
present the same scaling for the different order parameters. The
rotation $Q_{\mathrm{[010]_{pc}}}$ about [010]$_{\mathrm{pc}}$
reveals a scaling factor of 21.1~cm$^{-1}$/deg whereas the slope of
the rotation $Q_{\mathrm{[101]_{pc}}}$ about
[101]$_{\mathrm{pc}}$ gives 23.9~cm$^{-1}$/deg. This is natural and
expected when bearing in mind that the two soft modes are associated
to two independent order-parameters. The relation in
Eq.~\ref{relation} needs to be separately considered for each of the
relevant order parameters (in-phase and antiphase FeO$_{6}$ rotations
in our case), and there is no reason to expect that the values of the
coefficients in our Landau potential will be the same for
different cases. However, this difference was never pointed out in
previous investigations \cite{Iliev2006a, Chaix-Pluchery2011,
  Weber2012, Todorov2012}. This probably comes from a combination of
factors including experimental difficulties in mode assignment and
frequency determination, scattered data from a more limited number of
compounds, and possibly differences in scaling factors coincidentally
too small to be resolved experimentally. We believe that a careful
(re)investigation of the other series will reveal this
difference.

Last, we investigate the case of LaFeO$_3$ in more details. For
LaFeO$_3$, no $A_g$ Raman mode actually follows the scaling given by
the other members of the series. Instead, the $A_g$(2) and $A_g$(3)
modes, plotted as open triangles in Fig.~\ref{fig:TiltFrequ}, fall
below and above the scaling line respectively. On the other hand, we
have already pointed out that the band positions in LaFeO$_3$ differ
significantly from the other orthoferrites and do not seem to
follow from a continuous evolution of the other spectra. In order to
rationalize this comparatively exotic behaviour, we analyzed in
details the vibrational patterns given from our first-principles
calculations for LaFeO$_3$ and SmFeO$_3$. This comparison reveals
several frequency ranges where the modes do not keep their atomic
displacement patterns from La to Sm, but instead exhibit mixed
characteristics, which is expected from mode coupling phenomena
between two modes of the same symmetry getting close to each other. In
LaFeO$_3$, the mode mixing occurs in the regions between 100 and
200~cm$^{-1}$ and 400 and 450~cm$^{-1}$ as indicated in
Fig.~\ref{fig:Allwave}. In particular, it strongly affects the lower
soft mode $A_g$(3) as it approaches the lower lying $A_g$(2). For
Sm$^{3+}$ and smaller cations, these two modes have very
distinguishable atomic displacement patterns, with the $A_g$(2) mode
being dominated by \textit{R}$^{3+}$ displacements while $A_g$(3) is
dominated by octahedral rotations. In contrast, in LaFeO$_3$, the two
modes have significant contributions from both La$^{3+}$ displacement
and octahedral rotations. It is therefore no longer possible to
identify any of them as the soft mode of interest associated to
octahedral tilts only. The soft-mode frequency for a hypothetical
unmixed-state would lie between the two positions. This in turns
enable us to understand why the Raman spectrum LaFeO$_3$ is
significantly different as a whole from the others members of the
series, since the mode coupling will affect band positions and
intensities. This behavior par excellence has been reported by Iliev
et al. in orthomanganites \cite{Iliev2006b}, was also found in
the (La,Sm)CrO$_3$ solid solutions \cite{Daniels2013} and is probably
a general phenomenon occurring in orthorhombic \textit{Pnma}
perovskites in the limit of small tilt angles, where distortions of
the octahedra have to be taken into account \cite{Zhou2005}.

\section{Conclusion}
We have presented a Raman scattering study of a series of
orthoferrites \textit{R}FeO$_3$ (\textit{R}~= La, Sm, Eu, Gd, Tb,
Dy). A symmetry assignment of the observed modes has been presented on
the basis of a single-crystal study of SmFeO$_3$ and DFT calculations, and by taking advantage of the continuous changes in
the Raman spectra across the whole \textit{R}FeO$_3$ series. This
careful assignment has allowed to relate most of the vibration modes
to their vibrational pattern and symmetries. Based on this, we can
follow the structural evolution across the series and we have namely
shown that the $A_g$(3) and $A_g$(5) modes are the soft-modes of $A_g$
symmetry which correspond to the octahedral-rotation order-parameters
$Q_{[010]_{\mathrm{pc}}}$ (in-phase octahedral tilts about the
  [010] pseudocubic axis) and $Q_{[101]_{\mathrm{pc}}}$
(antiphase octahedral tilts about pseudocubic [101]). In this
framework we have demonstrated the proportionality of soft-mode
frequency and order parameter. Furthermore we have shown that
for rare-earth orthoferrites (and similar series) the change of the
soft-mode frequency depends only on the size of the rare earth (for a
fixed temperature). This work provides reference data for structural
investigation of the orthoferrite \textit{R}FeO$_3$ family, and
will be helpful in further studies of phenomena in orthoferrites
including structural instabilities, possible ferroelectricity and
multiferroicity, and rare-earth magnetism at low temperature via
spin-phonon coupling.

\begin{acknowledgments}
The authors thank R. Haumont (Université Paris Sud), B. Dkhil (Universit\'e Paris Saclay) and W. Ren (Shanghai University), E. Queiros and P.B. Tavares (University of Tr\'as-os-Montes e Alto Douro), and M. Mihalik jr., M. Mihalik, and M. Zentkova (Slovak Academy of Sciences) for providing high-quality samples.  MCW, MG, HJZ, JI and JK acknowledge financial support from the Fond National de Recherche Luxembourg through a PEARL grant (FNR/P12/4853155/Kreisel). JMA and AA acknowledge for financial support through projects Norte-070124-FEDER-000070, UID/NAN/50024/2013, PTDC/FIS-NAN/0533/2012 and VEGA2/0132/16. RV thanks for financial support through the grant PD/BI/106014/2015 by FCT.
\end{acknowledgments}


\end{document}